\documentclass{article}

\usepackage{arxiv}

\usepackage[utf8]{inputenc} 
\usepackage[T1]{fontenc}    
\usepackage{hyperref}       
\usepackage{url}            
\usepackage{booktabs}       
\usepackage{amsfonts}       
\usepackage{nicefrac}       
\usepackage{microtype}      
\usepackage{lipsum}
\usepackage{graphicx}
\usepackage{algorithm}
\usepackage{algorithmic}
\usepackage{amsmath}
\usepackage{subcaption}
\usepackage{tabularx}
\usepackage[table]{xcolor}
\usepackage{caption}
\usepackage{bbm}
\usepackage{booktabs}
\usepackage{siunitx}
\usepackage{threeparttable}
\usepackage{siunitx,tabularx,booktabs,threeparttable}
\usepackage[authoryear]{natbib}
\usepackage[section]{placeins}
\usepackage{flafter}

\setkeys{Gin}{width=\linewidth, height=0.55\textheight, keepaspectratio}

\newcommand{\best}[1]{\textbf{#1}}

\makeatletter
\def\fps@figure{tbp}
\makeatother

\usepackage{float}

\title{A Simulation Framework for Studying Systemic Effects of Feedback Loops in Recommender Systems}

\date{}

\AtBeginDocument{%
  \addtocontents{toc}{}
  \pagestyle{plain}     
}

\author{
 Gabriele Barlacchi \\
  Scuola Normale Superiore\\
  Pisa, Italy \\
  \texttt{gabriele.barlacchi@sns.it} \\
   \And
 Margherita Lalli \\
  Scuola Normale Superiore\\
  Pisa, Italy \\
  \texttt{margherita.lalli@sns.it} \\
  \And
 Emanuele Ferragina \\
  Sciences Po\\
  Paris \\
  \texttt{emanuele.ferragina@sciencespo.fr} \\
  \And
 Fosca Giannotti \\
  Scuola Normale Superiore\\
  Pisa, Italy \\
  \texttt{fosca.giannotti@sns.it} \\
  \And
 Luca Pappalardo \\
  CNR and Scuola Normale Superiore\\
  Pisa, Italy \\
  \texttt{luca.pappalardo@isti-cnr.it} \\
}

\begin{document}
\maketitle

\begin{abstract}
Recommender systems continuously interact with users, creating feedback loops that shape both individual behavior and collective market dynamics. This paper introduces a simulation framework to model these loops in online retail environments, where recommenders are periodically retrained on evolving user–item interactions. Using the Amazon e-Commerce dataset, we analyze how different recommendation algorithms influence diversity, purchase concentration, and user homogenization over time. 
Results reveal a systematic trade-off: while the feedback loop increases individual diversity, it simultaneously reduces collective diversity and concentrates demand on a few popular items. 
Moreover, for some recommender systems, the feedback loop increases user homogenization over time, making user purchase profiles increasingly similar.
These findings underscore the need for recommender designs that balance personalization with long-term diversity.
\end{abstract}


\section{Introduction}

Recommender systems constitute a pervasive layer of algorithmic mediation, increasingly shaping how individuals interact with digital environments.
On social media, they surface relevant posts and connections; in online retail, they suggest products that users are likely to purchase; and in location-based services, they propose efficient routes or destinations tailored to individual preferences.
Because recommender systems are powered by AI, their interaction with users naturally gives rise to a feedback loop: users’ choices shape the data on which recommenders are trained, and the trained models in turn influence subsequent user decisions \cite{pedreschi2025human}. 
This recursive process continually feeds back into itself, creating an evolving and potentially unbounded cycle. 

The feedback loop can amplify existing biases and generate various systemic effects, many of which are unintended \cite{pedreschi2025human, AI_SURVEY_PAPPALARDO}.  
On social media, the feedback loop may reinforce echo chambers, filter bubbles, and even processes of radicalization \cite{guess2023filterbubble, huszar2022algorithmic, sirbu2019algorithmic}; in online retail and location-based platforms, it can exacerbate popularity bias \cite{lee2018cross, aridor2024informational, MAURO_FEEDBACKLOOP, sanchez2023bias}.

Empirically studying feedback loops is challenging as experiments on real online platforms are limited by platform restrictions, short observation horizons, and poor reproducibility \cite{holzmeister2024replicability}. Consequently, most research relies on simulations, which enable controlled exploration of mechanisms and counterfactual scenarios \cite{general_FL_1, conterfactual_simulation,ALGOCONFOUNDING}. 
However, existing simulation approaches often rely on oversimplified assumptions, such as the exclusive use of explicit feedback (e.g., user ratings) \cite{FERRERO_EFFECTS}, the absence of periodic retraining of the recommender system \cite{cai2025agenticfeedbackloopmodeling}, and the consideration of only a limited set of recommendation algorithms \cite{debiasing_collabfilter}.

The difficulty of empirically studying and simulating feedback loops has limited our understanding of their systemic effects. In online retail, for instance, recommender systems are optimized for predictive accuracy and tend to favor popular items that are more likely to be purchased. While this typically increase overall sales volume from the platform’s perspective \cite{AI_SURVEY_PAPPALARDO, sales_boost}, it may also induce behavioral homogenization -- users increasingly buying the same items -- and concentrate demand on a small subset of products. Over time, these effects can distort online retail ecosystems and risk undermining market fairness and long-term system resilience. Yet, evidence on how the feedback loop leads to these systemic effects remains sparse and often contradictory \cite{AI_SURVEY_PAPPALARDO, fairness_contrary, hosanagar_2009}.

In this paper, we propose an open-source simulation framework that models feedback loops in an online-retail-like environment, where recommender systems are periodically retrained on the evolving user–item interaction data generated through purchases. 
The framework generates synthetic trajectories of user–item interactions over time, enabling detailed analysis of the systemic effects that emerge from the feedback loop. 
Unlike existing frameworks based on explicit feedback (e.g., user ratings), our framework uses implicit feedback (e.g., purchases), better reflecting real online retail operations and the recent shift toward implicit feedback modeling \cite{aggarwal2016recommender}.
The simulation framework is flexible and supports different recommendation systems to allow systematic comparison across algorithms.

We implement a wide set of recommendation systems and conduct extensive experiments using the open Amazon e-commerce 1.0 dataset \cite{AMAZON_10} to examine how the simulated feedback loop shapes systemic properties such as purchase diversity,  the concentration of purchase volume and item popularity across items, and behavioural homogenization across the user base. 
We discover that:

\begin{itemize}
\item The feedback loop broadens individual purchase profiles while concentrating demand at the collective level.
\item The exploratory boost induced by recommender systems is driven primarily by medium and heavy buyers, with minimal impact on light buyers.
\item The feedback loop generally increases the concentration of purchases on a small set of items across most recommender systems.
\item User homogenization effects are model-dependent: some recommender systems amplify behavioral similarity, while others preserve heterogeneity.
\end{itemize}

Our results demonstrate the utility of simulations to disentangle the systemic impacts of recommendations,  and highlight the need to design algorithms that balance personalization with unintended long-term effects.

\section{Related work} \label{section: related}
This section reviews research on how recommender systems influence user behavior, primarily in online retail, media, and news. 
The literature employs two approaches \cite{AI_SURVEY_PAPPALARDO}: \emph{online experiments} analyse real-world user interactions under controlled conditions and offer ecological validity, but face constraints from platform policies, short observation windows, and limited scalability; \emph{simulations} model user–system dynamics to explore hypothetical effects and counterfactual scenarios, but rely on assumptions that may limit external validity.

\paragraph{Online Experiments}
Empirical findings on consumption diversity are often conflicting. 
On the one hand, recommender systems can expand individual exploration but simultaneously concentrate demand at the aggregate level \cite{fleder2009blockbuster,lee2018cross}. On the other hand, recommendations may narrow user choices, steering individuals toward more homogeneous item sets \cite{aridor2024informational,wang2024field} and reduce listening diversity over time \cite{anderson2020spotify,holtz2020engagement}.
Research has also identified mechanisms through which recommendations influence decisions, including effects on item discovery, purchase behavior \cite{hinz2021causal}, and quality perceptions \cite{aridor2024informational} with broader implications for willingness to pay and market concentration \cite{goldfarb2011online,aral2012information,elberse2008longtail}.

Interface design moderates these effects, as shown in movie recommendation studies \cite{sun2024interactive}. 
News recommendation research highlights conditional effects: algorithmic exposure can amplify polarization \cite{levy2021polarization} or reinforce filter bubbles depending on user predispositions and design choices \cite{knudsen2023news,loecherbach2021diverse,haim2018filterbubble}. 
Diversity-aware frameworks have been proposed to address democratic concerns \cite{helberger2021fairness,moeller2021diversity}.
Even large-scale experiments reach opposite conclusions: some find no short-term filter-bubble effects on political attitudes \cite{guess2023filterbubble} while others document increased exposure to extreme content through recommendation chains \cite{ribeiro2020radicalization}.

\paragraph{Simulations} 
By isolating mechanisms and testing counterfactuals, simulations reveal feedback dynamics unfolding over extended periods.
Early work demonstrates that algorithmic reinforcement narrows exposure and reduces system-level diversity \cite{jiang2019degenerate} and repeated recommendations lead to behavioral homogenization. 
Mansoury et al. \cite{mansoury2020feedback} formalize how sustained exposure shapes user preferences rather than merely reflecting them. Agent-based models \cite{chaney2018algorithmic,mansoury2020popularity} reveal how small initial biases propagate into large-scale imbalances through mutual user-algorithm adaptation.
Beyond digital contexts, recent work \cite{zeng2021recwalk,MAURO_FEEDBACKLOOP} show that recommendation feedback loops extend into urban spatial behaviour.

Simulations also serve as testbeds for interventions. Helberger et al. \cite{helberger2019diversity} propose frameworks to balance diversity and autonomy in news, and Bauer et al. \cite{bauer2021diversity} investigate alternative strategies for music recommendation, and \cite{ALGO_DRIFT} proposes a framework for systematically evaluating recommender systems, with a particular focus on measuring drift in behavioral patterns and biases.
Interface design studies demonstrate how presentation shapes exposure  \cite{2022carousel}. Diversity-aware ranking and nudging strategies have been tested in simulation before deployment \cite{hazrati2021bias,mattis2022bias}.

Bias amplification under feedback loops has received particular attention. 
Hazrati and Ricci \cite{hazrati2021bias}  show how user biases distort evaluation metrics, with Mattis et al. \cite{mattis2022bias} demonstrating worsening effects over time. To address these risks, Hansen et al. \cite{hansen2023feedback} introduce debiasing techniques, and Aridor et al. \cite{aridor2024fairness} examine fairness-aware strategies accounting for systemic disparities.

Finally, modeling choices critically shape simulation outcomes. 
Zeng et al. \cite{zeng2020behavioral} demonstrate that behavioral model selection (e.g., multinomial logit vs. bounded rationality) significantly alters diversity predictions. While \cite{manco_kdd} introduce a flexible framework for generating synthetic preference data to support recommendation analysis, showing the differences among all the possible approaches.

Collectively, these contributions establish simulation as a bridge between theory and practice, enabling study of long-term dynamics hidden in short-term evaluations.

\section{Modelling of the Feedback Loop } \label{section: FL}
We consider a setting in which a recommender system on an online retail platform assists users by suggesting relevant items derived from historical interaction data. In this section, we first introduce the notation used throughout the paper (Section~\ref{sec:definitions}), then describe the simulation of the user–recommender feedback loop (Section~\ref{sec:framework}), and finally present the user choice model adopted in our framework (Section~\ref{sec:user_choice_model}).

\subsection{Definitions} \label{sec:definitions}
Let $\mathcal{U}$ be a set of users, $\mathcal{I}$ a set of items and $\mathcal{T}=\{1, 2 ..., T\}$ a sequence of discrete, ordered time steps.

We define a \textit{model scoring function} $R: \mathcal{U} \times \mathcal{I} \times \mathcal{T}    \rightarrow \mathbb{R}$ that assigns a relevance score $R(u,i,t)$ to every user-item-time triplet $ (u,i,t) \in \mathcal{U} \times \mathcal{I} \times \mathcal{T}$. 
This score is then used by a \textit{recommender function} $\rho: \mathcal{U} \times \mathcal{T} \rightarrow \mathcal{P}_k(\mathcal{I})$, where $\mathcal{P}_k(\mathcal{I})$ denotes the set of all ordered $k$-subsets of $\mathcal{I}$. 
For a given user $u$ and time $t$, this function generates a ranked list of $k$ items, denoted $K_{u,t}$, by selecting the items with highest scores: $ \rho(u, t) := K_{u,t} = \underset{i \in \mathcal{I}}{\mathrm{top\text{-}k}} \, R(u,i,t) $.

Users can also make purchases autonomously, independent of the recommender's suggestions. For these autonomous choices, a user's selection is constrained not by the full catalog $\mathcal{I}$ but by a user- and time-specific \textit{candidate set}, denoted $\mathcal{C}_{u,t}$. This set represents the subset of items that user $u$ is aware of at time $t$ and constitutes the available choice pool for organic interactions.

The \textit{user scoring function} is then a function $\hat{R}$ whose domain is the set of all admissible user-item-time triplets where the item is available to the user: $\hat{R}: \{ (u,i,t) \in \mathcal{U} \times \mathcal{I} \times \mathcal{T} \mid i \in \mathcal{C}_{u,t}  \} \rightarrow \mathbb{R}$. To each triplet $(u,i,t)$ in this domain, the   function assigns a score $\hat{R}(u,i,t)$.

\subsection{Simulation framework}
\label{sec:framework}
The simulation begins with an initialization phase, in which a real-world dataset is used to train both the model scoring function and the user scoring function.

The simulation is initialized at time $t_0$ and proceeds by generating user choices at each subsequent time step $t \in [t_0, T]$.

At each simulation step $t\in[t_0, T]$, a subset of users $\mathcal{U}_{t} \subseteq \mathcal{U}$ is awakened. Each active user $ u \in \mathcal{U}_{t}$ selects a basket of items, where the number of items selected, denoted by $b_{u,t}$, is the user's basket size for that step.

Each user-item interaction is modelled as follows:
\begin{itemize}
\item[\textbf{1.}] With probability $\eta$, the user $u$ selects an item from the ranked list $K_{u,t}$ provided by the recommender. 
This item is drawn from $K_{u,t}$ with a probability proportional to the scores $\{ R(u,i,t) \}_{i \in K_{u,t}}$.
\item[\textbf{2.}] With probability $1-\eta$, $u$ selects an item $i$ from the candidate set $\mathcal{C}_{u,t}$, according to a probability proportional to the score $\hat{R}(u,i,t)$ computed from the user choice model (described in the next section).
\end{itemize}

We partition the simulation timeline $[t_0, T]$ into discrete, non-overlapping intervals termed \emph{epochs}. Each epoch  $ \Delta_j$ (with $j$ ranging from $1$ to $n$) 
comprises a sequence of simulation steps, where the number of steps may vary between epochs.
At the end of each epoch $\Delta_j$, the model scoring function, the choice model, the candidate set and the ranked list suggested by the recommender are updated.

Algorithm~\ref{alg:cap} outlines the overall simulation framework, while Algorithm~\ref{alg2:cap} details the modeling of users’ next-item selection.

\begin{algorithm}
\caption{Simulation framework}\label{alg:cap}
\begin{algorithmic}
\REQUIRE Historical data $D$; total epochs $n$; last initialization epoch $E_{last\_init}$; retraining interval $\Delta$; cold-start $t_{0}$ 
;
\STATE \textbf{Output Data:} Interaction dataset $D_{post}$ 
\STATE $D_{init} \gets ColdStart(D, t_{0})$ 
\STATE $R\gets AlgorithmTraining(D_{init})$
\STATE $\hat{R} \gets ChoiceModelSetUp(D_{init})$
\STATE $E_{c} \gets E_{last\_init} + 1$
\STATE $E_{last\_training} \gets 0$
\STATE $D_{post} \gets D_{init}$
\WHILE{$E_{c} < E$}
\STATE $S_{e} \gets get\_steps\_epoch(E_c)$
\FORALL{$s$ in $S_{e}$}
\STATE $U_s \in \mathcal{U} \gets get\_awaken\_users(E_{c},s)$ 
\FORALL{$u$ in $S_{e}$}
\STATE $b_{u} \gets get\_basket(u)$
\STATE $i_{u_\text{next}} \gets ItemSelection(u, b_{u}, R, \hat{R})$
\STATE $D_{post} \gets D_{post} \cup \{(u, i_{u_\text{next}}, s)\}$
\ENDFOR
\IF{$E_c - E_{\text{last\_training}} > \Delta$}
\STATE $R \gets AlgorithmTraining(D_{post})$
\STATE $ \hat{R} \gets ChoiceModelSetUp(D_{post})$
\ENDIF
\ENDFOR
\ENDWHILE
\end{algorithmic}
\end{algorithm}

\begin{algorithm}
\caption{$Item selection$ for user $u$}\label{alg2:cap}
\begin{algorithmic}
\REQUIRE User $u$, basket size $b_u$, Choice model $\hat{R}$, Recommender model $R$, Current epoch $n$
\STATE \textbf{Hyperparameters:} Adoption rate $\eta$, Top ranking $K$
\STATE \textbf{Output Data:} Items list $U_I$
\STATE $CS \gets get\_candidate\_set(u, n)$
\FORALL{$i$ in $b_u$}
\IF{$Bernoulli(\eta)$}
\STATE $i\_next \gets R(u, K)$
\ELSE
\STATE $i\_next \gets \hat{R}(u, CS)$
\ENDIF
\ENDFOR
\end{algorithmic}
\end{algorithm}

\subsection{User choice model}
\label{sec:user_choice_model}

We model autonomous user choices -- i.e., the items selected when users do not follow the recommender system’s suggestions -- following the approach of Cesa-Bianchi et al. \cite{BOLTZMAN} and its subsequent extensions \cite{general_FL_1, ALGOCONFOUNDING, feedback_sim_ricci}. This model builds on item response theory and employs softmax decision rules, which are widely used in machine learning and discrete choice theory.

The probability that user $u$ selects an item $i$ from its candidate set $\mathcal{C}_{u,t}$ at time $t$ is given by a softmax function. We accomodate this by specifying the form of the scoring function $\hat{R}$ in terms of a utility function $V_{u,i}(t)$:
\begin{equation} \label{eq: choice prob}
P(i \mid u, t) = \frac{\hat{R}(u,i,t)}{\sum_{j \in \mathcal{C}_{u,t}} \hat{R}(u,j,t)} = \frac{e^{V_{u,i}(t)/\tau}}{\sum_{j \in \mathcal{C}_{u,t}} e^{V_{u,j}(t)/\tau}},
\end{equation}
where $V_{u,i}(t)$ is the utility associated with item  $i$ for user $u$ at time $t$ and $\tau$ controls the exploration–exploitation balance.
As $\tau \rightarrow 0$, choices become deterministic toward the highest-utility item, while for $\tau \rightarrow \infty$, they approach a uniform distribution.

\paragraph{Candidate set}

The set of available items for each user is constructed at each time by combining items from three distinct sources:
\begin{itemize}
    \item $GPop$: Items ranked by global popularity (i.e., the number of interactions by all users in the current epoch);
    \item $IPop$: Items ranked by individual popularity (i.e, the number of interactions from users' own purchase history up to the current epoch);
    \item $Unknown$: A random sample of items with which the user has had no prior interaction.
\end{itemize}

The final candidate set combines these sources with fixed proportions: 40\% from $GPop$, 40\% from $IPop$, and 20\% from $Unknown$. This composition balances globally and individually popular items with exploration of unseen content.

\paragraph{Utility estimation} \label{section:utility estimation}
We adapt the approach by Mansoury et al. \cite{general_FL_1} and estimate the utility of item $i$ for user $u$ as:
\begin{equation}
V_{u,i}(t) = c_u (t) + G_u(t) \cdot \log(1 + s_i(t)) + \lambda \cdot \frac{1}{1 + s_i(t)} + \eta_{u,i}(t)
\end{equation}
where $c_u$ denotes the average number of interactions of user $u$. The term $G_u \cdot \log(1 + s_i)$ modulates the influence of item popularity $s_i$ based on the user’s interaction diversity $G_u$, computed as the Gini index over the user's interaction distribution.  

The term $\lambda \cdot \frac{1}{1 + s_i}$ introduces a controlled boost for rare items, with $\lambda$ regulating the strength of this rarity effect \cite{HAZRATI, RARITY_1}. Finally, $\eta_{u,i}$ is a stochastic noise term, sampled at each timestep from a standard normal distribution. 
All the involved quantities ($c_u$, $G_u$, $s_i$) are evaluated by considering all simulation steps prior to $t$.

Such choice model serves as a structured null model that introduces a controlled degree of variability; more realistic than a simple popularity-based baseline, yet less arbitrary than random selection.

\section{Measuring systemic effects} \label{section: Metric}
We quantify the systemic effects of the feedback loop along three key dimensions: \emph{(i)} purchase diversity at individual and collective levels; \emph{(ii)} concentration of purchase volume and item popularity on a limited set of items; and \emph{(iii)} homogenization of user purchasing behaviour.

We select these measures due to the sparse and often conflicting evidence reported in the literature \cite{AI_SURVEY_PAPPALARDO} and because they operationalize fundamental dimensions of diversity and concentration in algorithmic marketplaces. In the long run, reduced collective diversity, excessive homogenization, and high concentration of purchases can limit item discoverability, reinforce popularity biases, and amplify inequality among items and sellers. 
Such dynamics may ultimately hinder market fairness, and reduce long-term system resilience.

\paragraph{Individual diversity}
We start by defining, for each user $u \in \mathcal{U}$, the purchase vector  $$\mathbf{w}_u = \{ w_{u,i} \mid i \in \mathcal{I}_u \},$$ where $w_{u,i}$ is the number of times $u$ purchased item $i$, and $\mathcal{I}_u$ the set of items in the purchase history of $u$.

We then use the Gini coefficient to quantify the inequality of weights in $\mathbf{w}_u$:
\begin{equation}\label{eq:Gini}
    G_u = \frac{\sum_{i=1}^{d_u} \sum_{j=1}^{d_u} |w_{ui} - w_{uj}|}{2 d_u^2 \overline{\mathbf{w}}_u}
\end{equation}
\noindent where $d_u = |\mathcal{I}_u|$ is the number distinct items $u$ has interacted with and $\overline{\mathbf{w}}_u$ is the average number of purchases made by $u$.

The set of all individual Gini coefficients $\{G_u\}_{u\in\mathcal{U}} $ is then summarised by the mean $\overline{G}_{ind}$ of the distribution.

\paragraph{Collective diversity}
We define a system-level vector $\mathbf{s}$, where each component corresponds to the total purchase volume of a specific item:
\begin{equation}\label{eq: prod strength}
\mathbf{s}  = \{ s_{i} \mid i \in \mathcal{I} \}, \quad \text{where} \; s_i = \sum_{u\in\mathcal{U}} w_{u,i}.
\end{equation}
where $s_{i}$ is denoted by strength of item $i$.
The collective Gini coefficient, $G_{\text{coll}}$, is then computed using Equation~\ref{eq:Gini} 
by replacing the individual weight vector $w_{u,i}$, the number of items $d_u$, and the mean weight $\overline{\mathbf{w}}_{u}$ with their system-level counterparts: the vector $\mathbf{s}$, the total number of items  $|\mathcal{I}|$, and the mean purchase volume per item $\overline{\mathbf{s}}$.

\paragraph{Purchase concentration}
To analyze the redistribution of purchases as the feedback loop evolves, we examine the frequency-rank curves of items at the end of the simulation using two key metrics: \emph{(ii)} the item purchase volume $s_i$ (or strength) defined in Eq. \eqref{eq: prod strength}; and \emph{(ii)} item popularity $p_i$, i.e., the number of unique users who purchased item $i$. 
This is given by $p_i = \sum_{u \in \mathcal{U}} \mathbbm{1}_{\{w_{u,i} > 0\}}$, where \( \mathbbm{1}_{\{w_{u,i} > 0\}} \) is an indicator function that equals 1 if user \( u \) has purchased item \( i \) at least once, and 0 otherwise.

\paragraph{User similarity}
We quantify the similarity between the interaction sets of two users $u, v \in \mathcal{U}$ with the Jaccard index: 

\begin{equation}
    J(u,v) = \frac{|\mathcal{I}_u \cap \mathcal{I}_v|}{|\mathcal{I}_u \cup \mathcal{I}_v|}
\end{equation}

The Jaccard index ranges from $0$, indicating no common items, to $1$, indicating identical item sets.
To obtain a system-level measure, we compute the mean of the Jaccard index across all unique user pairs as:
\begin{equation*}
    \bar{J}(\mathcal{U}) = \frac{1}{|\mathcal{P}(\mathcal{U})|}\sum_{(u,v) \in \mathcal{P}(\mathcal{U})} J(u,v)
\end{equation*}
where $\mathcal{P}(\mathcal{U})$ is the set of unordered distinct user pairs.

Items are then ranked by descending order based on $s_i$ and $p_i$, respectively, generating two independent ordered sequences $\{i_r\}_{r\in[1,|\mathcal{I}|]}$ such that $i_1$ has the highest value and  $i_{|\mathcal{I}|}$ the lowest. 

\section{Experiments} \label{section: experiments}

In this section, we describe the dataset, the recommender systems and their evaluation criteria, and the key parameters governing the simulation process.

\paragraph{Dataset}\label{Dataset}

We use the Amazon e-commerce 1.0 dataset \cite{berke2024open}, a large-scale collection of purchase histories crowdsourced from over 5,000 consumers between 2018 and 2022. It contains 1.8 million purchase events, each including detailed attributes such as item code, category, title, price, quantity, and the buyer’s shipping state, alongside anonymized user demographic information. 

To ensure temporal continuity during the simulation, we retain only users with at least one purchase per month over the entire simulated period. After filtering, the dataset comprises roughly one million interactions across 1,561 categories, generated by 2,191 users over a five-year period.

\paragraph{Recommender Systems}\label{section: recommenders}
We benchmark several recommender systems spanning classical collaborative filtering, matrix factorization, deep learning, and graph-based models (all implemented via the RecBole library \cite{recbole}):
\begin{itemize}
\item \textbf{ItemKNN} \cite{ITEM_KNN}: Item-based collaborative filtering that recommends items similar to those previously interacted with, using item–item co-occurrence similarity.
\item \textbf{MultiVAE} \cite{liang2018variational}: Variational autoencoder model that captures user–item interactions through latent variable inference.
\item \textbf{NeuMF} \cite{NEU_MF}: Neural matrix factorization approach that ranks items for each user by optimizing a pairwise probabilistic loss.
\item \textbf{LightGCN} \cite{LIGHT_GCN}: Simplified graph convolutional model that propagates neighborhood information without feature transformation or nonlinearities.
\item \textbf{BPR} \cite{BPR}: Matrix factorization model trained with a Bayesian pairwise ranking loss to learn personalized item rankings.
\item \textbf{SpectralCF} \cite{SPECTRAL}: Spectral graph-based collaborative filtering leveraging the frequency domain of the user–item graph to capture global connectivity.
\item \textbf{MostPop}: Non-personalized baseline recommending globally most popular items.
\end{itemize}

\paragraph{Evaluation of recommender systems}
We use the first six months of the dataset as an initialization period to calibrate the user choice model and to define the training, validation, and test splits.
Within this initialization phase, 80\% of the available time is assigned to training (months 1--4), 10\% to validation (month 5), and the remaining 10\% to testing (month 6).
This setup corresponds to a temporal holdout with a sliding window strategy, a widely adopted protocol to evaluate recommender systems over time \cite{timeholdout, jiang2022temporal}.

After initialization, we repeat the procedure iteratively: at the end of each subsequent epoch (one month of simulated data), we retrain models on the most recent four months, validate on the following month, and test on the newly generated interactions.
The configuration that maximized nDCG@10 on the validation set is selected.

For hyperparameter optimization, we employ exhaustive grid search over model-specific parameter ranges including learning rate, neighborhood size, and regularization terms.
As shown in Table \ref{tab:rec-results}, NeuMF achieves the best overall results across all metrics, followed by BPR. NeuMF outperforms the popularity-based baseline (MostPop) by +42\% in nDCG@10, +35\% in Recall@10, and +28\% in Precision@10. MostPop performs slightly better than ItemKNN, suggesting that naive similarity-based methods struggle under temporal dynamics. 

\begin{table}[t]
  \centering
  \caption{Performance comparison of recommendation systems.
  Best results are in \textbf{bold}.}
  \label{tab:rec-results}
  \begin{threeparttable}
    \begin{tabularx}{\columnwidth}{
        >{\raggedright\arraybackslash}X
        S[table-format=1.4]
        S[table-format=1.4]
        S[table-format=1.4]
        S[table-format=1.4]
    }
      \toprule
      Model & {N@10} & {P@10} & {R@10} & {Hit@10} \\
      \midrule
      MostPop    & 0.1398 & 0.0782 & 0.1706 & 0.5357 \\
      \midrule
      \textbf{Neu MF} & \best{0.2810} & \best{0.1398} & \best{0.2825} & \best{0.6699} \\
      BPR          & 0.2371 & 0.1072 & 0.2305 & 0.6241 \\
      Spectral CF  & 0.2274 & 0.1067 & 0.2292 & 0.6241 \\
      LightGCN     & 0.1969 & 0.0941 & 0.2033 & 0.5964 \\
      MultiVAE     & 0.1726 & 0.0934 & 0.2010 & 0.5942 \\
      Item-KNN     & 0.1539 & 0.0898 & 0.1945 & 0.5729 \\
      \bottomrule
    \end{tabularx}
    \begin{tablenotes}[para]
      \footnotesize
      N@10 = NDCG@10, P@10 = Precision@10, R@10 = Recall@10.
      All metrics are averaged over users; higher is better.
    \end{tablenotes}
  \end{threeparttable}
\end{table}

\paragraph{Simulation}
The Amazon dataset provides timestamps for each user-item interaction, enabling us to model simulation steps and epochs on a real temporal scale. 
In our setup, a single step  $t$ corresponds to one day and an epoch $\Delta_j$ corresponds to one month.
The initial part of the data is kept as an initialisation period and is used to train the recommender system and inform the user choice model. This period amounts to 6 months. Afterwards, our simulation carries on for $T=731$ days ($n=24$ epochs).

At each simulation step $t\in [t_0, T]$ ($t_0$ is the first day \textit{after} the initialization period), the set of active customers $\mathcal{U}_t$ and related basket sizes $\{b_{u,t}\}_{u\in \mathcal{U}_t}$ mimic the empirical purchase sequence of the corresponding day.
Consistent with typical streaming platform interfaces (10-20 items), the size of the ranked list $K_{u,t}$ is fixed to $k=20$ for each user and each time.

To evaluate the effects on the purchase dynamics of the recommendation systems described, we run the simulation in parallel for each model for different values of the adoption rate ($\eta$) values: $\{0, 0.2, 0.4, 0.6, 0.8, 1\}$. 
The case $\eta= 0$ corresponds to the evolution of the pure user choice model without influence of the recommender, while $\eta= 1$ represents the scenario in which users always accept the suggested recommendations.

To account for stochastic variability, we execute each configuration over three independent runs. 

\section{Results}
\label{sec:results}

\paragraph{Individual vs Collective Diversity} 
We compute the individual and collective Gini coefficients at the end of the 24-epoch period and evaluate how these values change as the adoption rate $\eta$ increases. 

\begin{figure}[!tbp]
    \centering
    \includegraphics[width=0.8\linewidth]{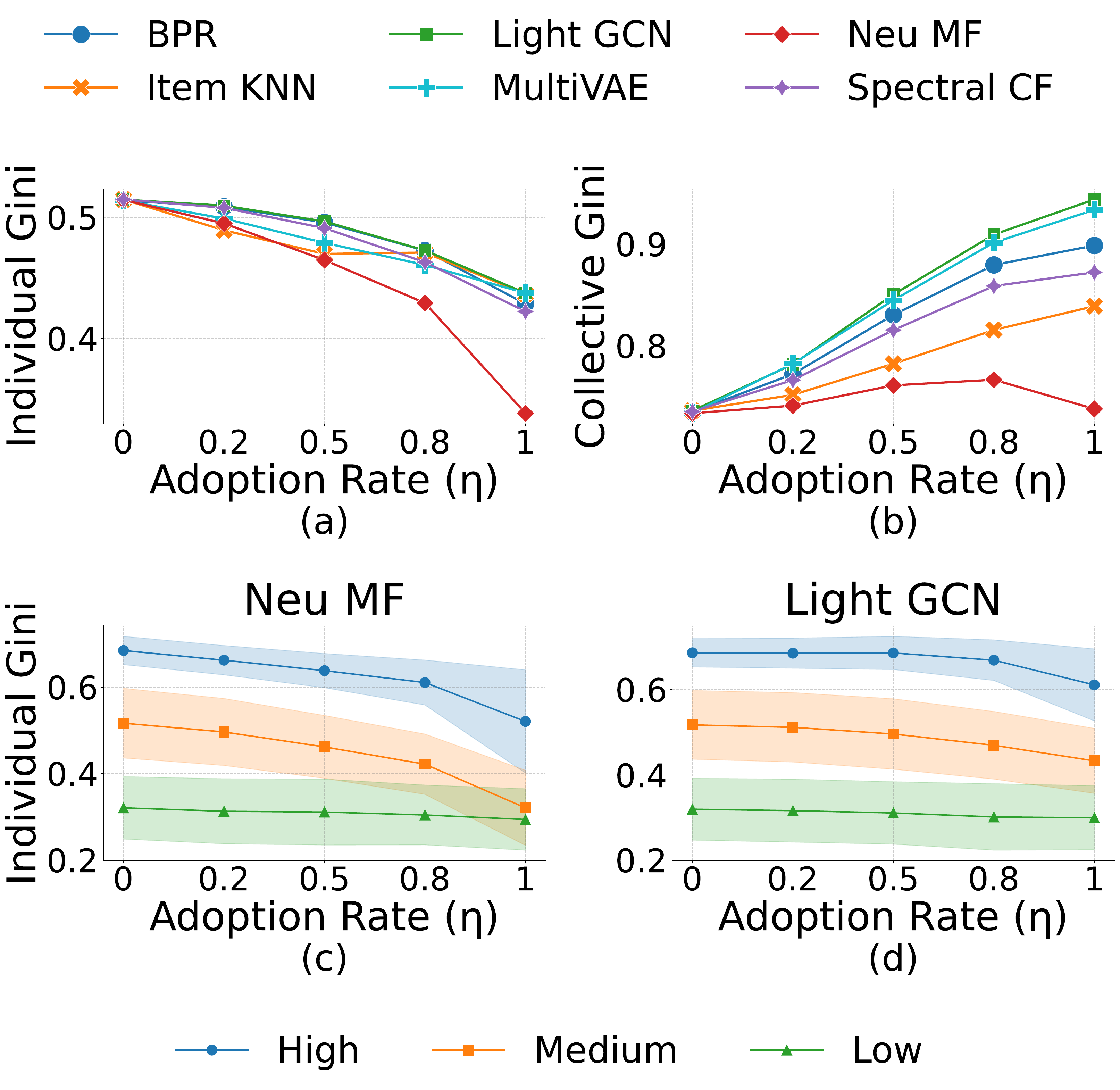}
    \caption{\textbf{Individual and collective diversity.} The evolution of individual (a) and collective (b) Gini coefficients as a function of adoption rate. Below, individual Gini coefficient disaggregated by user engagement segments for two representative models: NeuMF (c) and LightGCN (d). }
    \label{fig1:indVScoll diversity}
\end{figure}

We find a marked dichotomy in the response of individual versus collective diversity to recommender influence. 
As Figure \ref{fig1:indVScoll diversity}a shows, the average individual Gini coefficient exhibits a monotonic decrease as a function of $\eta$. 
The magnitude of this reduction ranges from approximately $11\%$ for LightGCN to $35\%$ for NeuMF, indicating a consistent enhancement in the diversity of individual purchase sequences.

In contrast, the collective-level Gini coefficient (Figure \ref{fig1:indVScoll diversity}b) demonstrates a positive correlation with $\eta$ for most models.
The magnitude of decrease in collective diversity reaches 28\% for LightGCN.
Interestingly, the most accurate recommender system -- and the one that most enhances individual exploration, NeuMF -- shows an initial drop in collective diversity, which later rises when $\eta = 1$, returning to levels similar to those observed at $\eta = 0$.  
Aside from this case, all other recommender systems combine increased individual exploration with decreased collective diversity. 

For a more granular analysis, Figure~\ref{fig1:indVScoll diversity}(c, d) disaggregates the average individual Gini coefficient by engagement segments, defined according to purchasing volume in the training set: heavy buyers (top 10\% of users, i.e., 10th decile), medium buyers (middle 80\%), and light buyers (bottom 10\%, i.e., 1st decile). We focus on the two models with the most distinct performance profiles, NeuMF and LightGCN. 

We find that the recommender system’s ability to drive diversity varies substantially across user groups.
The observed increase in individual diversity is predominantly attributable to the medium and heavy buyer segments, while the purchase diversity of light buyers remains nearly invariant with respect to $\eta$ for both models (see Figure~\ref{fig1:indVScoll diversity} (c, d)). This suggests that for low-engagement users, the models lack the necessary resolution in their learned representations to provide effective recommendations, thereby limiting their impact.

\paragraph{Concentration of purchases}

We compare the frequency-rank curves derived from the training set with those obtained from the simulation data, which includes the training set plus the subsequent 24 epochs of simulated purchases.

\begin{figure}[!tbp]
    \centering
    \includegraphics[width=0.8\linewidth,height=0.45\textheight,keepaspectratio]{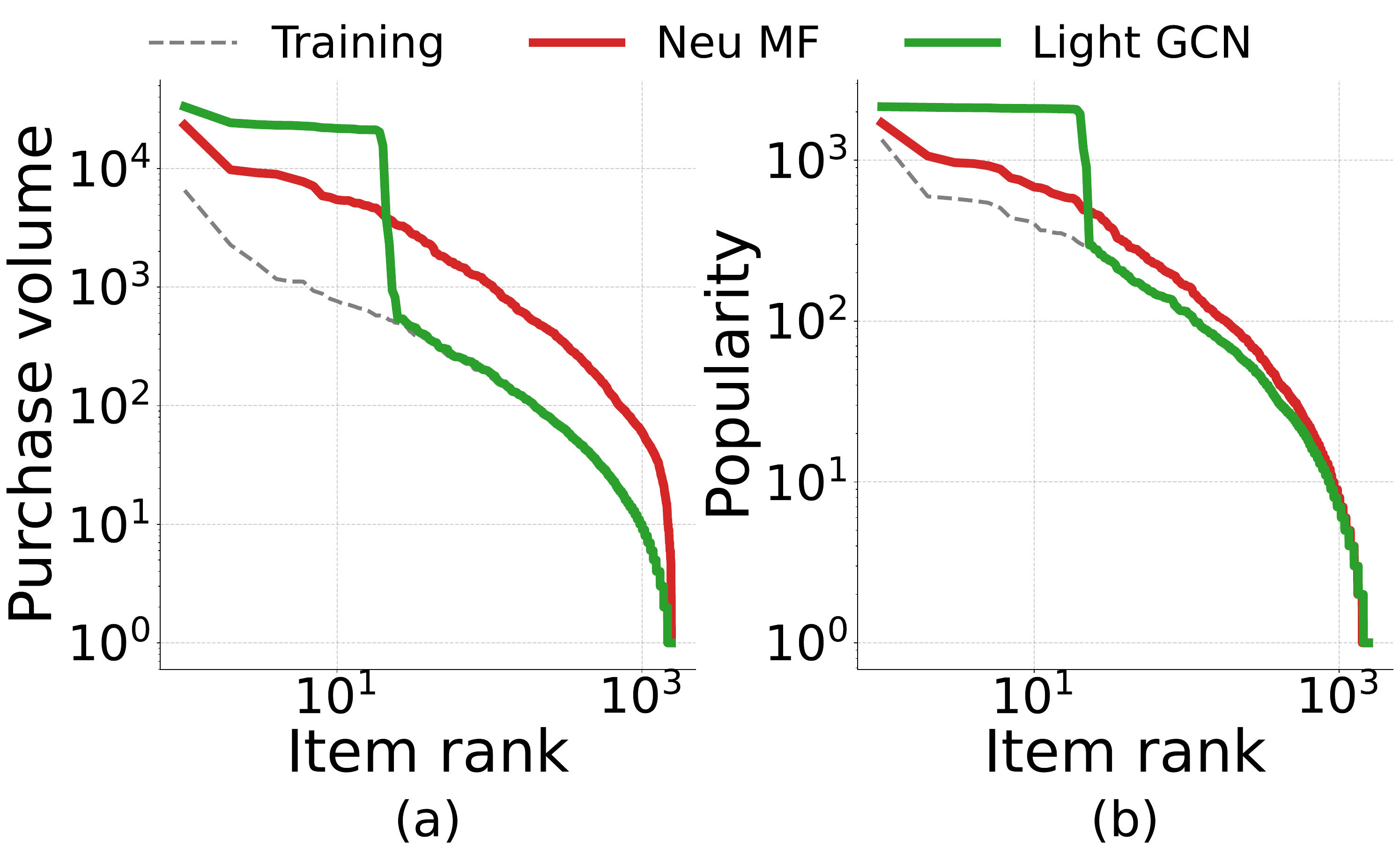}
    \caption{\textbf{Purchase redistribution for $\eta=1$}. Frequency-rank curves for the (a) purchase volume and (b) item popularity. Curves show the initial state (training set, grey) and the final state after 24 simulation epochs for NeuMF (green) and LightGCN (red).}
    \label{fig:Zipf}
\end{figure}

Such curves are presented in Figure~\ref{fig:Zipf}, with panel (a) corresponding to purchase volume and panel (b) to item popularity.
We focus on the two recommender systems that exhibit the most divergent behaviours for $\eta = 1$ (NeuMF and LightGCN) to make the effects more pronounced.

We find that LightGCN populates its recommendation set exclusively with the top 20 most popular items, leading to a stark drop-off in both the purchase volume and the popularity for items beyond the 20th rank (see green curves in Figure \ref{fig:Zipf}). In practice, both the purchase volume and the number of buyers for items ranked 21st and onwards remain stuck to their initial values in the training set (dashed curve in Figure \ref{fig:Zipf}). 

The purchase dynamics driven by the NeuMF recommendations reveal a more nuanced pattern:  the purchase volume distribution is overall unaffected \textit{in shape} (consistently with nearly invariant collective diversity for this recommender system), while the popularity distribution undergoes a concentration of mass, gaining a sharper peak and a heavier tail (see red curves in Figure~\ref{fig:Zipf}).
In particular, NeuMF stimulates increased purchasing across the entire spectrum of items, resulting in a quasi-uniform boost to purchase volumes.

A key insight emerges from comparison with the popularity-rank curve: while niche, lower-ranked items see an increase in total purchase volume, they fail to attract a broader customer base (see Figure~\ref{fig:Zipf}b). 
This is evidenced by the tail of the item popularity curve collapsing onto that of the gray baseline, indicating that there is no significant gain in the number of unique purchasers along the simulation. 
Consequently, the observed growth in volume for these niche items is primarily driven by existing users who had previously interacted with them. This stands in stark contrast to medium and top-selling items, which experience a concurrent boost in both purchase volume and popularity.

\begin{figure}
    \centering
    \includegraphics[width=0.8\linewidth]{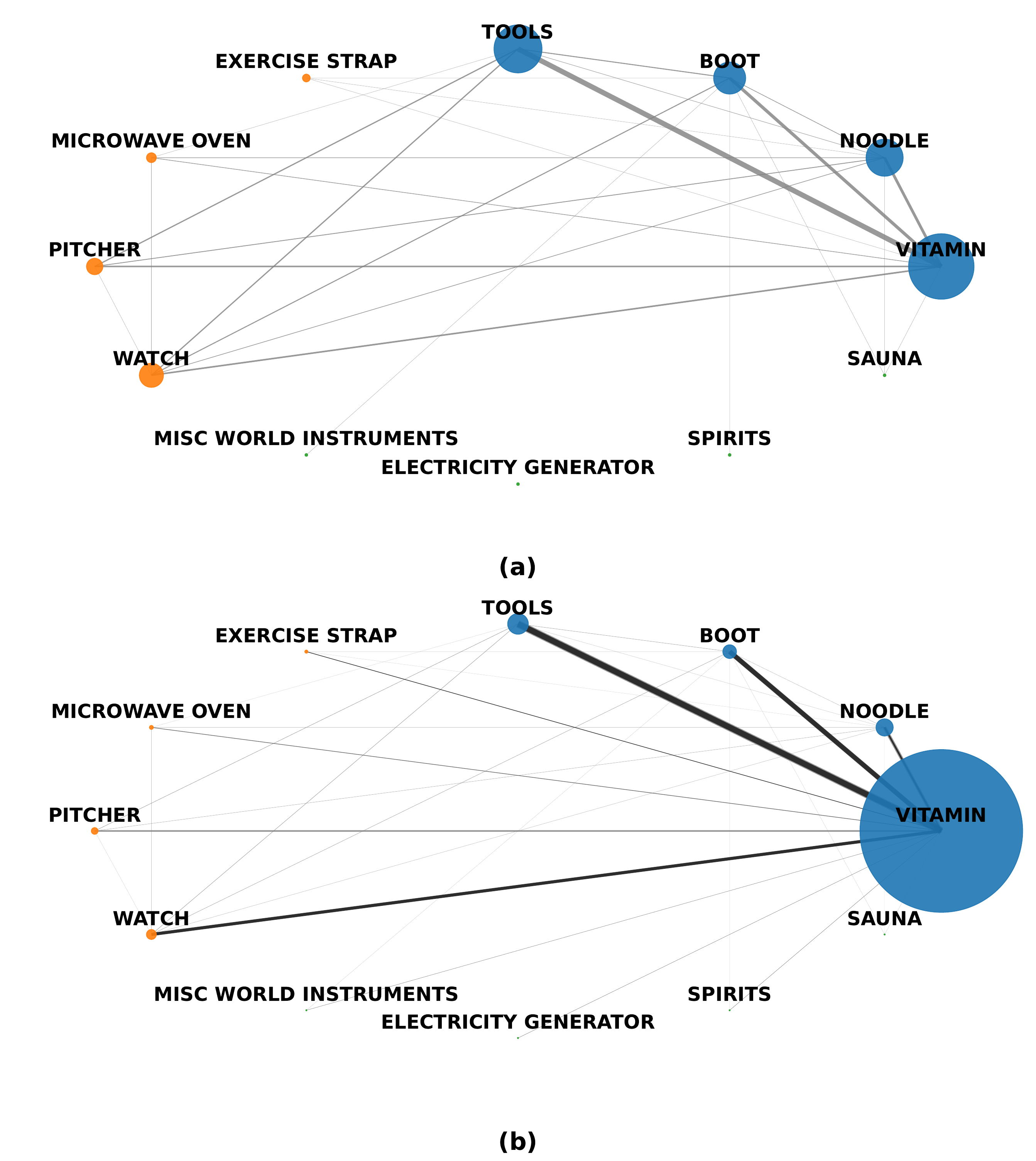}
    \caption{\textbf{Co-purchase networks before and after the feedback loop unfolds (LightGCN, $\eta = 1$, 24 epochs).}
    Each network includes a sample of items from different popularity levels. 
    Nodes represent items (sized by total purchases), and edges connect items with shared buyers (thicker edges mean more shared purchasers).
    }
    \label{fig:networks}
\end{figure}

To illustrate an example of the effect of increasing concentration, Figure~\ref{fig:networks} compares the co-purchase networks -- where nodes represent item categories and weighted edges indicate how frequently two categories are purchased together -- for LightGCN at epoch~0 and epoch~24.
At epoch~0, the network displays a relatively balanced topology, with low variance in node degrees and edge weights.
After 24 epochs, a pronounced core-periphery structure emerges, characterized by a dominant hub of popular items (e.g., vitamins). This indicates a clear systemic concentration of purchases as the feedback loop unfolds for 24 epochs. 

\paragraph{User homogenization} 

\begin{figure}[!tbp]
    \centering
    \includegraphics[width=0.8\linewidth]{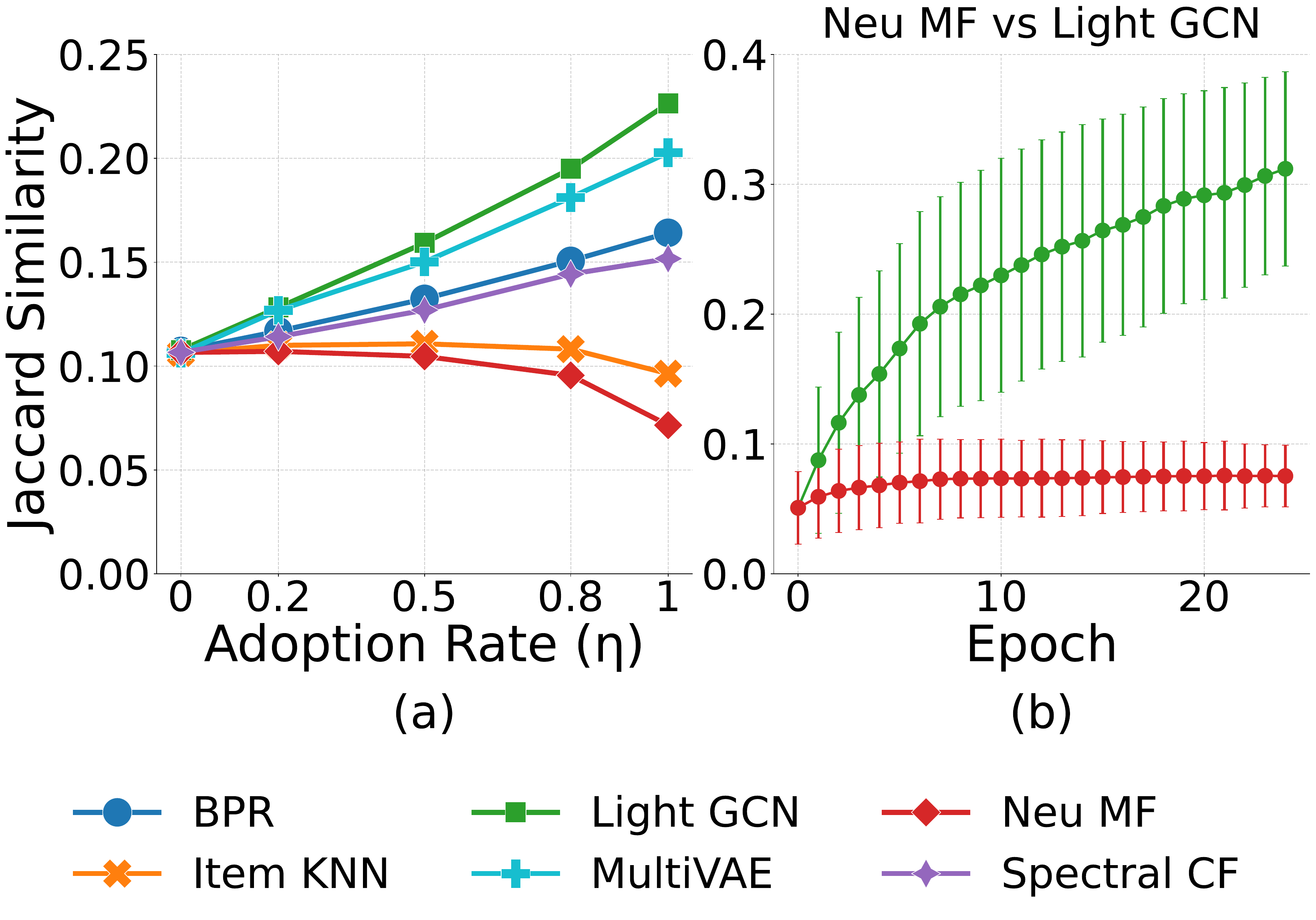}
    \caption{\textbf{User homogenization.} (a) User similarity as a function of adoption rate ($\eta$): LightGCN and MultiVAE strongly increase homogenization, BPR and Spectral CF show moderate growth, while NeuMF and Item-KNN reduce it at high adoption. (b) For ($\eta=0.8$), NeuMF remains stable, while LightGCN steadily amplifies user similarity over time.}
    \label{fig1:jaccard}
\end{figure}

Figure \ref{fig1:jaccard}a shows the evolution of user similarity as a function of $\eta$ across different recommendation systems. 

We find that LightGCN and MultiVAE exerts the strongest homogenizing force: as adoption rate approaches $\eta = 1$, the average similarity across users increases sharply, reaching a maximum increase of 110\% for LightGCN.
This indicates that the recommender systems consistently push individuals toward the same popular items. 
BPR and SpectralCF also induce convergence, though to a lesser degree, suggesting that their latent representations capture a broader but still overlapping range of preferences. 
In contrast, Item-KNN and NeuMF produce the opposite (even if slight) tendency: as adoption intensifies, the similarity curve declines, with a maximum reduction of 33\% for NeuMF. 
This implies that when users rely entirely on recommendations from these systems, their trajectories remain more differentiated.

Figure \ref{fig1:jaccard}b reports the evolution of user similarity over the 24 epochs for a fixed adoption rate ($\eta = 0.8$), contrasting the two most divergent models: NeuMF and LightGCN. 
NeuMF exhibits a flat trajectory where user similarity remains stable over time: it preserves heterogeneity across users. 
In contrast, LightGCN shows a steady upward trend: epoch after epoch, user choices grow more aligned, underscoring its strong tendency to channel demand toward the same set of items.
LightGCN disrupt heterogeneity across users.

Overall, these findings reveal that homogenization is model-dependent. Recommender systems such as LightGCN or MultiVAE, foster behavioral convergence by over-amplifying structural popularity signals. Systems like NeuMF, which balance linear and nonlinear components of user–item interactions, better preserve heterogeneity even under high recommendation adoption. This divergence has critical implications: while all recommenders affect diversity, only certain models systematically erode distinctiveness in user behavior, thus magnifying risks of cultural uniformity and reduced exposure to niche content.

\FloatBarrier
\section{Discussion and Conclusions} \label{section: conclusions}
This paper introduced a simulation framework to model the feedback loop of recommender systems in online retail environments and investigate its systemic effects.

We find a crucial trade-off in individual vs collective diversity: as recommendation adoption increases, individual diversity increases while collective diversity declines sharply.
Only NeuMF -- a neural model achieving the highest performance among those tested -- preserves collective diversity at levels comparable to those observed in the training data.
These results confirm the empirical findings of Lee and Hosanagar \cite{lee2018cross}, who showed through controlled empirical experiments on an online retail platform that collaborative filtering increases individual diversity while reducing collective diversity. 
Our study extends their work by showing that this pattern holds across a wide range of recommender systems and becomes more pronounced as the adoption rate of the recommender increases.

We also observe an increasing concentration of purchase volume and item popularity as the feedback loop unfolds: the apparent rise in individual diversity primarily translates into repeated purchases of a few already popular items. This pattern mirrors recent findings on feedback loops in location-based recommender systems \cite{MAURO_FEEDBACKLOOP}, suggesting that such systemic effects may emerge universally across different types of online platforms.

Finally, we find that user homogenization effects are model-dependent: some recommender systems strongly align user purchase behaviors, while others introduce a modest increase in behavioral heterogeneity. This suggests that certain systemic patterns arise inherently from the recursive nature of the feedback loop, whereas others are shaped by the specific training dynamics of each recommender model.

Our study has some limitations. First, although the choice model accounts for repeated consumption and evolving awareness, it abstracts away important behavioral factors such as price sensitivity, brand loyalty, and novelty-seeking -- all of which are critical in retail settings. Future work should incorporate these dimensions to improve ecological validity. Second, while the framework is grounded in large-scale purchase histories, its simulation-based nature calls for external validation using longitudinal observational or experimental data to assess generalizability. Finally, the frequency of recommender retraining is also likely to influence systemic outcomes, a factor we plan to investigate in future work.
A promising research direction emerging from our study concerns the design of interventions within the feedback loop to mitigate the observed systemic effects. For instance, re-ranking strategies could be introduced to progressively enhance collective diversity while reducing user homogenization and purchase concentration over time.

In conclusion, our simulation framework opens new avenues for studying the coevolution between algorithmic decision support
and user choices on online retail platform, contributing to a deeper understanding of how digital
platforms shape, and are shaped by, human behaviour.

\newpage
\bibliographystyle{plainnat}
\bibliography{reference}

\begin{thebibliography}{66}
\providecommand{\natexlab}[1]{#1}
\providecommand{\url}[1]{\texttt{#1}}
\expandafter\ifx\csname urlstyle\endcsname\relax
  \providecommand{\doi}[1]{doi: #1}\else
  \providecommand{\doi}{doi: \begingroup \urlstyle{rm}\Url}\fi

\bibitem[A.~Berke and Mahari(2024)]{AMAZON_10}
D.~Calacci A.~Berke and R.~Mahari.
\newblock Open e-commerce 1.0, five years of crowdsourced u.s. amazon purchase histories with user demographics, 2024.
\newblock URL \url{https://doi.org/10.1038/s41597-024-03329-6}.

\bibitem[Aggarwal(2016)]{aggarwal2016recommender}
Charu~C. Aggarwal.
\newblock \emph{Recommender Systems: The Textbook}.
\newblock Springer Publishing Company, Incorporated, 1st edition, 2016.
\newblock ISBN 3319296574.

\bibitem[Anderson et~al.(2020)Anderson, Maystre, Anderson, Mehrotra, and Lalmas]{anderson2020spotify}
Ashton Anderson, Lucas Maystre, Ian Anderson, Rishabh Mehrotra, and Mounia Lalmas.
\newblock Algorithmic effects on the diversity of consumption on spotify.
\newblock In \emph{Proceedings of The Web Conference 2020 (WWW '20)}, pages 2155--2165. ACM, 2020.

\bibitem[Aral et~al.(2012)Aral, Brynjolfsson, and Alstyne]{aral2012information}
Sinan Aral, Erik Brynjolfsson, and Marshall~Van Alstyne.
\newblock Information, technology, and information worker productivity.
\newblock \emph{Information Systems Research}, 23\penalty0 (3):\penalty0 849--867, 2012.

\bibitem[Aridor et~al.(2024{\natexlab{a}})Aridor, Goncalves, Kluver, Kong, and Konstan]{aridor2024informational}
Guy Aridor, Duarte Goncalves, Daniel Kluver, Ruoyan Kong, and Joseph Konstan.
\newblock The informational role of online recommendations: Evidence from a field experiment, 2024{\natexlab{a}}.
\newblock Working Paper.

\bibitem[Aridor et~al.(2024{\natexlab{b}})Aridor, Kong, and Konstan]{aridor2024fairness}
Guy Aridor, Ruoyan Kong, and Joseph~A. Konstan.
\newblock Fairness-aware recommendation through simulation-based evaluation.
\newblock In \emph{Proceedings of the 18th ACM Conference on Recommender Systems (RecSys ’24)}, 2024{\natexlab{b}}.

\bibitem[Bauer and Jannach(2021)]{bauer2021diversity}
Christine Bauer and Dietmar Jannach.
\newblock Balancing diversity and relevance in music recommendation.
\newblock In \emph{Proceedings of the 15th ACM Conference on Recommender Systems (RecSys ’21)}, pages 179--188, 2021.

\bibitem[Berke et~al.(2024)Berke, Calacci, Mahari, Yabe, Larson, and Pentland]{berke2024open}
Alex Berke, Dan Calacci, Robert Mahari, Takahiro Yabe, Kent Larson, and Sandy Pentland.
\newblock Open e-commerce 1.0, five years of crowdsourced us amazon purchase histories with user demographics.
\newblock \emph{Scientific Data}, 11\penalty0 (1):\penalty0 491, 2024.

\bibitem[Cai et~al.(2025)Cai, Zhang, Bao, Gao, Wang, Feng, and He]{cai2025agenticfeedbackloopmodeling}
Shihao Cai, Jizhi Zhang, Keqin Bao, Chongming Gao, Qifan Wang, Fuli Feng, and Xiangnan He.
\newblock Agentic feedback loop modeling improves recommendation and user simulation, 2025.
\newblock URL \url{https://arxiv.org/abs/2410.20027}.

\bibitem[Campos et~al.(2010)Campos, D{\'\i}ez, and Cantador]{timeholdout}
Pedro~G Campos, Fernando D{\'\i}ez, and Iv{\'a}n Cantador.
\newblock Temporal evaluation of recommender systems on historical data.
\newblock In \emph{Proceedings of the 4th ACM Conference on Recommender Systems}, pages 81--88, 2010.

\bibitem[Cesa-Bianchi et~al.(2017)Cesa-Bianchi, Gentile, Lugosi, and Neu]{BOLTZMAN}
Nicolò Cesa-Bianchi, Claudio Gentile, Gábor Lugosi, and Gergely Neu.
\newblock Boltzmann exploration done right, 2017.
\newblock URL \url{https://arxiv.org/abs/1705.10257}.

\bibitem[Chaney et~al.(2018{\natexlab{a}})Chaney, Stewart, and Engelhardt]{ALGOCONFOUNDING}
Allison J.~B. Chaney, Brandon~M. Stewart, and Barbara~E. Engelhardt.
\newblock How algorithmic confounding in recommendation systems increases homogeneity and decreases utility.
\newblock In \emph{Proceedings of the 12th ACM Conference on Recommender Systems}, page 224–232. ACM, September 2018{\natexlab{a}}.
\newblock \doi{10.1145/3240323.3240370}.
\newblock URL \url{http://dx.doi.org/10.1145/3240323.3240370}.

\bibitem[Chaney et~al.(2018{\natexlab{b}})Chaney, Stewart, and Engelhardt]{chaney2018algorithmic}
Allison J.~B. Chaney, Brandon~M. Stewart, and Barbara~E. Engelhardt.
\newblock How algorithmic confounding in recommendation systems increases homogeneity and decreases utility.
\newblock In \emph{Proceedings of the 12th ACM Conference on Recommender Systems (RecSys ’18)}, pages 224--232, 2018{\natexlab{b}}.

\bibitem[Coppolillo et~al.(2024)Coppolillo, Mungari, Ritacco, Fabbri, Minici, Bonchi, and Manco]{ALGO_DRIFT}
Erica Coppolillo, Simone Mungari, Ettore Ritacco, Francesco Fabbri, Marco Minici, Francesco Bonchi, and Giuseppe Manco.
\newblock Algorithmic drift: A simulation framework to study the effects of recommender systems on user preferences, 2024.
\newblock URL \url{https://arxiv.org/abs/2409.16478}.

\bibitem[Elberse(2008)]{elberse2008longtail}
Anita Elberse.
\newblock Should you invest in the long tail?
\newblock \emph{Harvard Business Review}, 86\penalty0 (7/8):\penalty0 88--96, 2008.

\bibitem[Fleder and Hosanagar(2009)]{fleder2009blockbuster}
Daniel Fleder and Kartik Hosanagar.
\newblock Blockbuster culture's next rise or fall: The impact of recommender systems on sales diversity.
\newblock \emph{Management Science}, 55\penalty0 (5):\penalty0 697--712, 2009.

\bibitem[Fleder et~al.(2010)Fleder, Hosanagar, and Buja]{hosanagar_2009}
Daniel Fleder, Kartik Hosanagar, and Andreas Buja.
\newblock Recommender systems and their effects on consumers: the fragmentation debate.
\newblock In \emph{Proceedings of the 11th ACM Conference on Electronic Commerce}, EC '10, page 229–230, New York, NY, USA, 2010. Association for Computing Machinery.
\newblock ISBN 9781605588223.
\newblock \doi{10.1145/1807342.1807378}.
\newblock URL \url{https://doi.org/10.1145/1807342.1807378}.

\bibitem[Goldfarb and Tucker(2011)]{goldfarb2011online}
Avi Goldfarb and Catherine Tucker.
\newblock Online display advertising: Targeting and obtrusiveness.
\newblock \emph{Marketing Science}, 30\penalty0 (3):\penalty0 389--404, 2011.

\bibitem[Guess et~al.(2023)Guess, Nagler, and Tucker]{guess2023filterbubble}
Andrew Guess, Jonathan Nagler, and Joshua Tucker.
\newblock The filter bubble hypothesis: Experimental evidence.
\newblock \emph{Science Advances}, 9\penalty0 (14):\penalty0 eade0140, 2023.

\bibitem[Haim et~al.(2018)Haim, Graefe, and Brosius]{haim2018filterbubble}
Mario Haim, Andreas Graefe, and Hans-Bernd Brosius.
\newblock Burst of the filter bubble? effects of personalization on the diversity of google news.
\newblock \emph{Digital Journalism}, 6\penalty0 (3):\penalty0 330--343, 2018.

\bibitem[Hansen et~al.(2023)Hansen, Mehrotra, McInerney, and Biega]{hansen2023feedback}
Casper Hansen, Rishabh Mehrotra, James McInerney, and Asia~J. Biega.
\newblock Correcting feedback loops in recommendation with simulation-based debiasing.
\newblock In \emph{Proceedings of the 17th ACM Conference on Recommender Systems (RecSys ’23)}, pages 411--421, 2023.

\bibitem[Hazrati and Ricci(2021)]{hazrati2021bias}
Maryam Hazrati and Francesco Ricci.
\newblock Bias amplification in recommender systems: Simulation study.
\newblock In \emph{Proceedings of the 29th ACM Conference on User Modeling, Adaptation and Personalization (UMAP ’21)}, pages 240--244, 2021.

\bibitem[Hazrati and Ricci(2022)]{HAZRATI}
Naieme Hazrati and Francesco Ricci.
\newblock Recommender systems effect on the evolution of users’ choices distribution.
\newblock \emph{Information Processing and Management}, 59\penalty0 (1):\penalty0 102766, 2022.
\newblock ISSN 0306-4573.
\newblock \doi{https://doi.org/10.1016/j.ipm.2021.102766}.
\newblock URL \url{https://www.sciencedirect.com/science/article/pii/S0306457321002466}.

\bibitem[Hazrati et~al.(2020)Hazrati, Elahi, and Ricci]{feedback_sim_ricci}
Naieme Hazrati, Mehdi Elahi, and Francesco Ricci.
\newblock Simulating the impact of recommender systems on the evolution of collective users' choices.
\newblock In \emph{Proceedings of the 31st ACM Conference on Hypertext and Social Media}, HT '20, page 207–212, New York, NY, USA, 2020. Association for Computing Machinery.
\newblock ISBN 9781450370981.
\newblock \doi{10.1145/3372923.3404812}.
\newblock URL \url{https://doi.org/10.1145/3372923.3404812}.

\bibitem[He et~al.(2017)He, Liao, Zhang, Nie, Hu, and Chua]{NEU_MF}
Xiangnan He, Lizi Liao, Hanwang Zhang, Liqiang Nie, Xia Hu, and Tat-Seng Chua.
\newblock Neural collaborative filtering, 2017.
\newblock URL \url{https://arxiv.org/abs/1708.05031}.

\bibitem[He et~al.(2020)He, Deng, Wang, Li, Zhang, and Wang]{LIGHT_GCN}
Xiangnan He, Kuan Deng, Xiang Wang, Yan Li, Yongdong Zhang, and Meng Wang.
\newblock Lightgcn: Simplifying and powering graph convolution network for recommendation, 2020.
\newblock URL \url{https://arxiv.org/abs/2002.02126}.

\bibitem[Helberger et~al.(2019)Helberger, Karppinen, and D'Acunto]{helberger2019diversity}
Natali Helberger, Kari Karppinen, and Lucia D'Acunto.
\newblock On the democratic role of news recommenders.
\newblock \emph{Digital Journalism}, 7\penalty0 (8):\penalty0 993--1012, 2019.

\bibitem[Helberger et~al.(2021)Helberger, Trilling, Moeller, and Tekinerdogan]{helberger2021fairness}
Natali Helberger, Damian Trilling, Judith Moeller, and Bedir Tekinerdogan.
\newblock Fairness in online personalization: Perspectives from communication and journalism studies.
\newblock \emph{Digital Journalism}, 9\penalty0 (2):\penalty0 236--255, 2021.

\bibitem[Hinz et~al.(2021)Hinz, Li, and Grahl]{hinz2021causal}
Oliver Hinz, Xitong Li, and J{\"o}rn Grahl.
\newblock How do recommender systems lead to consumer purchases? a causal mediation analysis of a field experiment.
\newblock \emph{Information Systems Research}, 33\penalty0 (2):\penalty0 620--637, 2021.
\newblock \doi{10.1287/isre.2021.1074}.

\bibitem[Holtz et~al.(2020)Holtz, Carterette, Chandar, Nazari, Cramer, and Aral]{holtz2020engagement}
David Holtz, Ben Carterette, Praveen Chandar, Zahra Nazari, Henriette Cramer, and Sinan Aral.
\newblock The engagement-diversity connection: Evidence from a field experiment on spotify.
\newblock In \emph{Proceedings of the 21st ACM Conference on Economics and Computation (EC '20)}, pages 75--76. ACM, 2020.

\bibitem[Holzmeister et~al.(2024)]{holzmeister2024replicability}
Felix Holzmeister et~al.
\newblock Examining the replicability of online experiments.
\newblock \emph{Nature Human Behaviour}, 2024.
\newblock Demonstrates challenges and replication failures in web-based experiments due to platform and API limitations.

\bibitem[Husz{\'a}r et~al.(2022)Husz{\'a}r, Ktena, O’Brien, Belli, Schlaikjer, and Hardt]{huszar2022algorithmic}
Ferenc Husz{\'a}r, Sofia~Ira Ktena, Conor O’Brien, Luca Belli, Andrew Schlaikjer, and Moritz Hardt.
\newblock Algorithmic amplification of politics on twitter.
\newblock \emph{Proceedings of the national academy of sciences}, 119\penalty0 (1):\penalty0 e2025334119, 2022.

\bibitem[Jiang et~al.(2019)Jiang, Chiappa, Lattimore, Gy{\"o}rgy, and Kohli]{jiang2019degenerate}
Ray Jiang, Silvia Chiappa, Tor Lattimore, Andr{\'a}s Gy{\"o}rgy, and Pushmeet Kohli.
\newblock Degenerate feedback loops in recommender systems.
\newblock In \emph{Proceedings of the AAAI Conference on Artificial Intelligence}, volume~33, pages 4275--4282, 2019.

\bibitem[Jiang et~al.(2022)Jiang, Wang, Lee, Liu, Wang, and Ye]{jiang2022temporal}
Yanan Jiang, Hao Wang, Kuan-Chuan Lee, Jun Liu, Meng Wang, and Jieping Ye.
\newblock Temporal collaborative filtering with bayesian probabilistic tensor factorization.
\newblock \emph{ACM Transactions on Information Systems}, 40\penalty0 (4):\penalty0 1--30, 2022.

\bibitem[Knudsen(2023)]{knudsen2023news}
Erik Knudsen.
\newblock Modeling news recommender systems’ conditional effects on selective exposure: Evidence from two online experiments.
\newblock \emph{Journal of Communication}, 73\penalty0 (2):\penalty0 138--149, 2023.

\bibitem[Lee and Hosanagar(2018)]{lee2018cross}
Dokyun Lee and Kartik Hosanagar.
\newblock How do recommender systems affect sales diversity? a cross-category investigation via randomized field experiment.
\newblock \emph{Information Systems Research}, 30\penalty0 (1):\penalty0 239--259, 2018.

\bibitem[Levy(2021)]{levy2021polarization}
Ro’ee Levy.
\newblock Social media, news consumption, and polarization: Evidence from a field experiment.
\newblock \emph{American Economic Review}, 111\penalty0 (3):\penalty0 831--870, 2021.

\bibitem[Liang et~al.(2018)Liang, Krishnan, Hoffman, and Jebara]{liang2018variational}
Dawen Liang, Rahul~G Krishnan, Matthew~D Hoffman, and Tony Jebara.
\newblock Variational autoencoders for collaborative filtering.
\newblock In \emph{Proceedings of the 2018 World Wide Web Conference}, pages 689--698, 2018.

\bibitem[Loecherbach et~al.(2021)Loecherbach, Welbers, Moeller, Trilling, and Atteveldt]{loecherbach2021diverse}
Felicia Loecherbach, Kasper Welbers, Judith Moeller, Damian Trilling, and Wouter~Van Atteveldt.
\newblock Is this a click towards diversity? explaining when and why news users make diverse choices.
\newblock In \emph{Proceedings of the 13th ACM Web Science Conference 2021 (WebSci '21)}, pages 282--290. ACM, 2021.

\bibitem[Mansoury et~al.(2020{\natexlab{a}})Mansoury, Abdollahpouri, Pechenizkiy, Mobasher, and Burke]{general_FL_1}
Masoud Mansoury, Himan Abdollahpouri, Mykola Pechenizkiy, Bamshad Mobasher, and Robin Burke.
\newblock Feedback loop and bias amplification in recommender systems, 2020{\natexlab{a}}.
\newblock URL \url{https://arxiv.org/abs/2007.13019}.

\bibitem[Mansoury et~al.(2020{\natexlab{b}})Mansoury, Abdollahpouri, Pechenizkiy, Mobasher, and Burke]{mansoury2020feedback}
Masoud Mansoury, Himan Abdollahpouri, Mykola Pechenizkiy, Bamshad Mobasher, and Robin Burke.
\newblock Feedback loops in recommender systems: Quantification and mitigation.
\newblock In \emph{Proceedings of the 29th ACM International Conference on Information and Knowledge Management (CIKM ’20)}, pages 2145--2148, 2020{\natexlab{b}}.

\bibitem[Mansoury et~al.(2020{\natexlab{c}})Mansoury, Abdollahpouri, Pechenizkiy, Mobasher, and Burke]{mansoury2020popularity}
Masoud Mansoury, Himan Abdollahpouri, Mykola Pechenizkiy, Bamshad Mobasher, and Robin Burke.
\newblock Popularity bias in recommendation: A multi-stakeholder perspective.
\newblock In \emph{Proceedings of the 13th ACM Conference on Recommender Systems (RecSys ’20)}, pages 342--346, 2020{\natexlab{c}}.

\bibitem[Mattis et~al.(2022)Mattis, Burke, and Abdollahpouri]{mattis2022bias}
John Mattis, Robin Burke, and Himan Abdollahpouri.
\newblock Modeling user bias evolution in recommender systems.
\newblock In \emph{Proceedings of the 16th ACM Conference on Recommender Systems (RecSys ’22)}, pages 648--652, 2022.

\bibitem[Mauro et~al.(2025)Mauro, Minici, and Pappalardo]{MAURO_FEEDBACKLOOP}
Giovanni Mauro, Marco Minici, and Luca Pappalardo.
\newblock The urban impact of ai: Modeling feedback loops in next-venue recommendation, 2025.
\newblock URL \url{https://arxiv.org/abs/2504.07911}.

\bibitem[Moeller et~al.(2021)Moeller, Helberger, and Trilling]{moeller2021diversity}
Judith Moeller, Natali Helberger, and Damian Trilling.
\newblock Diversity by design: Investigating the potential of diversity-aware recommender systems in journalism.
\newblock \emph{Journalism Studies}, 22\penalty0 (11):\penalty0 1469--1488, 2021.

\bibitem[Mungari et~al.(2025)Mungari, Coppolillo, Ritacco, and Manco]{manco_kdd}
Simone Mungari, Erica Coppolillo, Ettore Ritacco, and Giuseppe Manco.
\newblock Flexible generation of preference data for recommendation analysis.
\newblock In \emph{Proceedings of the 31st ACM SIGKDD Conference on Knowledge Discovery and Data Mining V.2}, KDD '25, page 5710–5721, New York, NY, USA, 2025. Association for Computing Machinery.
\newblock ISBN 9798400714542.
\newblock \doi{10.1145/3711896.3737398}.
\newblock URL \url{https://doi.org/10.1145/3711896.3737398}.

\bibitem[Pappalardo et~al.(2024)Pappalardo, Ferragina, Citraro, Cornacchia, Nanni, Rossetti, Gezici, Giannotti, Lalli, Gambetta, et~al.]{AI_SURVEY_PAPPALARDO}
Luca Pappalardo, Emanuele Ferragina, Salvatore Citraro, Giuliano Cornacchia, Mirco Nanni, Giulio Rossetti, Gizem Gezici, Fosca Giannotti, Margherita Lalli, Daniele Gambetta, et~al.
\newblock A survey on the impact of ai-based recommenders on human behaviours: methodologies, outcomes and future directions.
\newblock \emph{arXiv preprint arXiv:2407.01630}, 2024.

\bibitem[PATHAK et~al.(2010)PATHAK, GARFINKEL, GOPAL, VENKATESAN, and YIN]{sales_boost}
BHAVIK PATHAK, ROBERT GARFINKEL, RAM~D. GOPAL, RAJKUMAR VENKATESAN, and FANG YIN.
\newblock Empirical analysis of the impact of recommender systems on sales.
\newblock \emph{Journal of Management Information Systems}, 27\penalty0 (2):\penalty0 159--188, 2010.
\newblock ISSN 07421222.
\newblock URL \url{http://www.jstor.org/stable/29780174}.

\bibitem[Pedreschi et~al.(2025)Pedreschi, Pappalardo, Ferragina, Baeza-Yates, Barab{\'a}si, Dignum, Dignum, Eliassi-Rad, Giannotti, Kert{\'e}sz, et~al.]{pedreschi2025human}
Dino Pedreschi, Luca Pappalardo, Emanuele Ferragina, Ricardo Baeza-Yates, Albert-L{\'a}szl{\'o} Barab{\'a}si, Frank Dignum, Virginia Dignum, Tina Eliassi-Rad, Fosca Giannotti, J{\'a}nos Kert{\'e}sz, et~al.
\newblock Human-ai coevolution.
\newblock \emph{Artificial Intelligence}, 339:\penalty0 104244, 2025.

\bibitem[Rahdari et~al.(2024)Rahdari, Brusilovsky, and Kveton]{2022carousel}
Behnam Rahdari, Peter Brusilovsky, and Branislav Kveton.
\newblock Towards simulation-based evaluation of recommender systems with carousel interfaces.
\newblock \emph{ACM Transactions on Recommender Systems}, 2024.
\newblock \doi{https://doi.org/10.1145/3643709}.
\newblock To appear / published online (depends on version).

\bibitem[Rendle et~al.(2009)Rendle, Freudenthaler, Gantner, and Schmidt-Thieme]{BPR}
Steffen Rendle, Christoph Freudenthaler, Zeno Gantner, and Lars Schmidt-Thieme.
\newblock Bpr: Bayesian personalized ranking from implicit feedback.
\newblock In \emph{Proceedings of the Twenty-Fifth Conference on Uncertainty in Artificial Intelligence}, UAI '09, page 452–461, Arlington, Virginia, USA, 2009. AUAI Press.
\newblock ISBN 9780974903958.

\bibitem[Ribeiro et~al.(2020)Ribeiro, Ottoni, West, Almeida, and Jr.]{ribeiro2020radicalization}
Manoel~Horta Ribeiro, Raphael Ottoni, Robert West, Virg{\'i}lio A.~F. Almeida, and Wagner~Meira Jr.
\newblock Auditing radicalization pathways on youtube.
\newblock In \emph{Proceedings of the 2020 ACM Conference on Fairness, Accountability, and Transparency (FAT* '20)}, pages 131--141. ACM, 2020.

\bibitem[S{\'a}nchez et~al.(2023)S{\'a}nchez, Bellog{\'\i}n, and Boratto]{sanchez2023bias}
Pablo S{\'a}nchez, Alejandro Bellog{\'\i}n, and Ludovico Boratto.
\newblock Bias characterization, assessment, and mitigation in location-based recommender systems.
\newblock \emph{Data Mining and Knowledge Discovery}, 37\penalty0 (5):\penalty0 1885--1929, 2023.

\bibitem[Sarwar et~al.(2001)Sarwar, Karypis, Konstan, and Riedl]{ITEM_KNN}
Badrul Sarwar, George Karypis, Joseph Konstan, and John Riedl.
\newblock Item-based collaborative filtering recommendation algorithms.
\newblock In \emph{Proceedings of the 10th International Conference on World Wide Web}, WWW '01, page 285–295, New York, NY, USA, 2001. Association for Computing Machinery.
\newblock ISBN 1581133480.
\newblock \doi{10.1145/371920.372071}.
\newblock URL \url{https://doi.org/10.1145/371920.372071}.

\bibitem[S{\^\i}rbu et~al.(2019)S{\^\i}rbu, Pedreschi, Giannotti, and Kert{\'e}sz]{sirbu2019algorithmic}
Alina S{\^\i}rbu, Dino Pedreschi, Fosca Giannotti, and J{\'a}nos Kert{\'e}sz.
\newblock Algorithmic bias amplifies opinion fragmentation and polarization: A bounded confidence model.
\newblock \emph{PloS one}, 14\penalty0 (3):\penalty0 e0213246, 2019.

\bibitem[Sonboli et~al.(2022)Sonboli, Burke, Ekstrand, and Mehrotra]{fairness_contrary}
Nasim Sonboli, Robin Burke, Michael Ekstrand, and Rishabh Mehrotra.
\newblock The multisided complexity of fairness in recommender systems.
\newblock \emph{AI Mag.}, 43\penalty0 (2):\penalty0 164–176, June 2022.
\newblock ISSN 0738-4602.
\newblock \doi{10.1002/aaai.12054}.
\newblock URL \url{https://doi.org/10.1002/aaai.12054}.

\bibitem[Sun et~al.(2024)Sun, Akella, Kong, Zhou, and Konstan]{sun2024interactive}
Ruixuan Sun, Avinash Akella, Ruoyan Kong, Moyan Zhou, and Joseph~A. Konstan.
\newblock Interactive content diversity and user exploration in online movie recommenders: A field experiment.
\newblock \emph{International Journal of Human–Computer Interaction}, 40\penalty0 (22):\penalty0 7233--7247, 2024.

\bibitem[Sun et~al.(2019)Sun, Khenissi, Nasraoui, and Shafto]{debiasing_collabfilter}
Wenlong Sun, Sami Khenissi, Olfa Nasraoui, and Patrick Shafto.
\newblock Debiasing the human-recommender system feedback loop in collaborative filtering.
\newblock In \emph{Companion Proceedings of The 2019 World Wide Web Conference}, WWW '19, page 645–651, New York, NY, USA, 2019. Association for Computing Machinery.
\newblock ISBN 9781450366755.
\newblock \doi{10.1145/3308560.3317303}.
\newblock URL \url{https://doi.org/10.1145/3308560.3317303}.

\bibitem[Wang et~al.(2024)Wang, Sun, Yuan, and Chen]{wang2024field}
Yitong Wang, Tianshu Sun, Zhe Yuan, and AJ~Yuan Chen.
\newblock How recommendation affects customer search: A field experiment.
\newblock \emph{Information Systems Research}, 36\penalty0 (1):\penalty0 85--106, 2024.

\bibitem[Yang et~al.(2021)Yang, Dai, Dong, Chen, He, and Wang]{conterfactual_simulation}
Mengyue Yang, Quanyu Dai, Zhenhua Dong, Xu~Chen, Xiuqiang He, and Jun Wang.
\newblock Top-n recommendation with counterfactual user preference simulation.
\newblock pages 2342--2351, 10 2021.
\newblock \doi{10.1145/3459637.3482305}.

\bibitem[Yao et~al.(2021)Yao, Halpern, Thain, Wang, Lee, Prost, Chi, Chen, and Beutel]{FERRERO_EFFECTS}
Sirui Yao, Yoni Halpern, Nithum Thain, Xuezhi Wang, Kang Lee, Flavien Prost, Ed~H. Chi, Jilin Chen, and Alex Beutel.
\newblock Measuring recommender system effects with simulated users, 2021.
\newblock URL \url{https://arxiv.org/abs/2101.04526}.

\bibitem[Zeng et~al.(2020)Zeng, Fang, and Si]{zeng2020behavioral}
Yan Zeng, Yuan Fang, and Luo Si.
\newblock User behavior modeling in recommender system simulations.
\newblock In \emph{Proceedings of the 43rd International ACM SIGIR Conference on Research and Development in Information Retrieval (SIGIR ’20)}, pages 1509--1512, 2020.

\bibitem[Zeng et~al.(2021)Zeng, Chen, Sun, Fang, and Si]{zeng2021recwalk}
Yan Zeng, Zhenzhong Chen, Jianyuan Sun, Yuan Fang, and Luo Si.
\newblock Recwalk: Nearly unbiased walk-based recommendation via markov chains.
\newblock In \emph{Proceedings of the 14th ACM International Conference on Web Search and Data Mining (WSDM ’21)}, pages 367--375, 2021.

\bibitem[Zhao et~al.(2021)Zhao, Mu, Hou, Lin, Chen, Pan, Li, Lu, Wang, Tian, Min, Feng, Fan, Chen, Wang, Ji, Li, Wang, and Wen]{recbole}
Wayne~Xin Zhao, Shanlei Mu, Yupeng Hou, Zihan Lin, Yushuo Chen, Xingyu Pan, Kaiyuan Li, Yujie Lu, Hui Wang, Changxin Tian, Yingqian Min, Zhichao Feng, Xinyan Fan, Xu~Chen, Pengfei Wang, Wendi Ji, Yaliang Li, Xiaoling Wang, and Ji-Rong Wen.
\newblock Recbole: Towards a unified, comprehensive and efficient framework for recommendation algorithms, 2021.
\newblock URL \url{https://arxiv.org/abs/2011.01731}.

\bibitem[Zheng et~al.(2018)Zheng, Lu, Jiang, Zhang, and Yu]{SPECTRAL}
Lei Zheng, Chun-Ta Lu, Fei Jiang, Jiawei Zhang, and Philip~S. Yu.
\newblock Spectral collaborative filtering.
\newblock In \emph{Proceedings of the 12th ACM Conference on Recommender Systems}, RecSys '18, page 311–319, New York, NY, USA, 2018. Association for Computing Machinery.
\newblock ISBN 9781450359016.
\newblock \doi{10.1145/3240323.3240343}.
\newblock URL \url{https://doi.org/10.1145/3240323.3240343}.

\bibitem[Zhou et~al.(2010)Zhou, Kuscsik, Liu, Medo, Wakeling, and Zhang]{RARITY_1}
Tao Zhou, Zoltán Kuscsik, Jian-Guo Liu, Matúš Medo, Joseph~Rushton Wakeling, and Yi-Cheng Zhang.
\newblock Solving the apparent diversity-accuracy dilemma of recommender systems.
\newblock \emph{Proceedings of the National Academy of Sciences}, 107\penalty0 (10):\penalty0 4511--4515, 2010.
\newblock \doi{10.1073/pnas.1000488107}.
\newblock URL \url{https://www.pnas.org/doi/abs/10.1073/pnas.1000488107}.

\end{thebibliography}

\end{document}